\documentclass[aps,pra,floatfix,superscriptaddress,preprint,showpacs]{revtex4}

\pdfoutput=1

\usepackage{graphicx}
\usepackage{graphics}
\usepackage{amssymb}
\usepackage{amsmath}
\usepackage{latexsym}
\usepackage{color}
\usepackage{iphysics}

\begin{document}

\title{The Jahn-Teller instability in dissipative quantum electromechanical systems}

\author{Charles P. Meaney, Tim Duty, Ross H. McKenzie, Gerard J. Milburn}
\affiliation{Department of Physics, School of Mathematical and Physical Sciences, The University of Queensland, St Lucia, QLD 4072, Australia}

\begin{abstract}
We consider the steady states of a harmonic oscillator coupled so strongly to a two-level system (a qubit) that the rotating wave approximation cannot be made. The Hamiltonian version of this model is known as the $E\otimes\beta$ Jahn-Teller model. The semiclassical version of this system exhibits a fixed point bifurcation, which in the quantum model leads to a ground state with substantial entanglement between the oscillator and the qubit. We show that the dynamical bifurcation survives in a dissipative quantum description of the system, amidst an even richer bifurcation structure.   We propose two experimental implementations of this model based on superconducting cavities:  a parametrically driven nonlinear nanomechanical resonator coupled capacitively to a coplanar microwave cavity and a superconducting junction in the central conductor of a coplanar waveguide. 
\end{abstract}
\pacs{85.85.+j; 42.50.Wk; 82.40.Bj; 84.40.Dc; 85.25.-j  }

\maketitle

\section{Introduction.}
Circuit quantum electrodynamics\cite{Devoret:2007} and quantum nanomechanics\cite{Schwab-Roukes}  have emerged in the last few years as new experimental contexts for the study of strongly coupled quantum systems. A superconducting coplanar microwave cavity  can support very strong electric fields with very low dissipation. The electric field due to a single photon can be as large as  $0.2\ $V/m and cavity quality factors as high as $10^6$  have been obtained\cite{Blais}. Large electric dipoles, in the form of superconducting junction elements, can be placed into the gap between the central conductor and coupled into the cavity field which is treated as a simple harmonic oscillator.  The coupling strength can now be made far larger than equivalent experiments in atomic physics, and may yet be made still larger\cite{Devoret:2007}. Alternatively, a nanomechanical resonator can form one plate of a capacitor coupling a coherent driving field to the cavity field\cite{Lehnert, Woolley1}. If the nanomechanical element exhibits a significant Duffing nonlinearity then under parametric driving it can be approximated as a two level system interacting very strongly with the cavity resonator\cite{Wielinga}.   In this paper we consider a model in which the coupling between the field and the two level system is so large that we cannot make the rotating wave approximation.   The model that results is known as the $E\tp \beta$ Jahn-Teller model.

The $E\tp \beta$ Jahn-Teller model describes the interaction between a single simple harmonic oscillator and a two level system, or qubit. It has recently received some attention as there is a critical coupling strength at which the nature of the ground state undergoes a morphological change reflecting a bifurcation in a fixed point of the corresponding classical model\cite{JT-bifur}. As the coupling strength increases the ground state entanglement increases monotonically. Starting from a zero coupling strength and increasing to one infinitely large, the ground states change from:
\beq
 \K{0}_c\K{0}_q \to \K{\al}_c\K{+}_q + \K{{-}\al}_c\K{-}_q
\eeq
where $\K{0}_c$ and $\K{0}_q$ are the bare cavity oscillator and qubit ground states and $\K{{\pm}\al}_c$ is an oscillator coherent state while $\K{\pm}_q$ are orthogonal qubit states. The rate of change of entanglement as a function of the coupling strength is greatest for coupling strengths near the critical coupling strength for a fixed point bifurcation in the corresponding semiclassical description\cite{JT-bifur}.  This has a significant implications for the ability to reach the zero photon state in the cavity by cooling.   If one were to engineer a system with a coupling strength above the critical value, cooling would reach a ground state in which the number of photons in the field was $\abs{\al}^2$ not zero, while the qubit would be measured to be in a totally mixed state for $\al$ only a little large than unity.   These statements of course apply only to Hamiltonian systems, without damping. As real systems have finite damping, no matter how small, it is our objective here to determine to what extent the ground state bifurcation exhibited in the conservative system is manifest in the damped system and further to specify experimental scenarios in circuit QED in which it may be observed.

The paper is structured as follows. In section I we present a detailed analysis of the fixed point structure of the dissipative Jahn-Teller $E\otimes\beta$ model in a semiclassical description. We include both dissipation of the oscillator and the two-level system. Surprisingly, despite the large number of parameters in the mode, the bifurcation structure is shown to depend on only three dimensionless independent parameters. In section II we consider the quantum version of the model in which dissipation is described using a Markov master equation of the oscillator and the two-level system. We numerically determine the steady state of the system. After tracing out the two-level system, we construct the Q-fuction for the oscillator. We then show that as the control parameters are varied through the values at which the semiclassical model shows bifurcations the Q-function changes from single peaked to double peaked with support on the semiclassical fixed points. In section III we present two physical systems in circuit quantum electrodynamics and quantum nanomechanics that     could be used to implement the dissipative Jahn-Teller model. We suggest a number of key experimental signatures of the bifurcation. Finally in section IV we summarise our results and suggest new directions for further work.

\section{The dissipative $E\tp \beta$ Jahn-Teller model. }
We consider the case of a two level system coupled to a simple harmonic oscillator. The coupling is linear in the oscillator displacement and represents a state dependent constant force acting on the oscillator. The two-level system Hamiltonian includes a term which mixes the eigenstates of the conditional displacement.   In order to model a realistic device we also need to include dissipation of both the oscillator and the two level system. The oscillator is damped into a zero temperature heat bath  with an amplitude decay rate of $\ka$. The two-level system is assumed to undergo spontaneous emission at rate $\ga$, and phase decay at rate $\iGa$. The irreversible dynamics due to these processes will be described by a master equation for weak damping (and with the rotating wave approximation for the system-bath coupling) , for density matrix, $\rho$,  of the total resonator and qubit system.
\bea
 \Dt{\dm} & =   {}-\frac{\ii}{\hb}\C{\ham}{\dm} + \ka\<{2\aop\dm\ad-\ada\dm-\dm\ada} \\
          & \es {}+\frac{\ga}{2}\<{2\te{\si_{-_x}\rh\si_{+_x}}-\te{\si_{+_x}\si_{-_x}\rh}-\te{\rh\si_{+_x}\si_{-_x}}} - \frac{\iGa}{8}\C{\six}{\C{\six}{\dm}}
\eea
where:
\bea
 \ham           & = \hb\om\ada + \frac{\hb\iDe}{2}\six + \frac{\hb\ep}{2}\siz + \hb\lambda\<{\aop+\ad}\siz + \hb\et\<{\aop+\ad}
 \label{JT-ham}
\eea
and:
\bea
 \te{\si_{\pm_x}} & = \frac{1}{2}\<{\siz\mp\ii\siy}
\eea
where $\sigma_x$ etc. are Pauli matrices. 
Note that we have modelled dissipation of the qubit as spontaneous emission in the eigenbasis of $\sigma_x$. This is based on the assumption that the free Hamiltonian of the qubit is simply proportional to $\sigma_x$. This makes our model consistent with the Hamiltonian model discussed in \cite{JT-bifur} which has no dissipation and $\epsilon=0$, and ensures that for this limit, the fixed points of the two models coincide. The coupling between the qubit and the oscillator is modelled by the term proportional to $\lambda$ and represents a state dependent force acting on the oscillator. Alternatively, we can think of this term as describing a dependance of the energy eigenstates of the qubit on an oscillator degree of freedom, as occurs in the orginal Jahn-Teller model in which the electronic energy levels are dependent on one or more relative nuclear coordinates.  Finally we have added an independent resonant force acting on the oscillator degree of freedom through the term proportional to $\eta$. In the circuit QED realisation this would represent a driving voltage applied to the co-planar cavity.

In the absence of dissipation, the semiclassical equations of motion that follow for the Hamiltonian have a fixed point pitch-fork bifurcation\cite{JT-bifur} at a critical coupling strength of 
$\lambda_{cr}=\frac{\sqrt{\iDe\om}}{2}$. A single stable elliptic fixed point, with zero cavity field amplitude, below the bifurcation changes stability to give two new elliptic fixed points with equal and opposite cavity field amplitude. When dissipation is included we expect that a similar bifurcation will occur but in this case the fixed points will be zero dimensional attractors. We can study this bifurcation by deriving the semiclassical equations of motion as follows. Using the master equation we construct moment equations for the expectation of each of the two-level system operators ($\six$, $\siy$, and $\siz$) and for the quadrature phase field operators defined by
\bea
 \te{X} & = \frac{1}{2}\<{\aop+\ad} \\
 \te{Y} & = -\ii\frac{1}{2}\<{\aop-\ad}
\eea
which satisfy the commutation relations, $\C{\te{X}}{\te{Y}}=\ii/2$.
The equations of motion for the expectations of the five quantities of interest are found to be:
\bea
 \Dt{\E{\te{X}}} & = \om\E{\te{Y}}-\ka\E{\te{X}} \\
 \Dt{\E{\te{Y}}} & = -\et-\la \E{\siz}-\om\E{\te{X}}-\ka\E{\te{Y}} \\
 \Dt{\E{\six}}   & = -\ep \E{\siy}-4\la \E{\te{X\si_y}}-\ga\<{1+\E{\six}} \\
 \Dt{\E{\siy}}   & = -\iDe \E{\siz}+\ep \E{\six}+4\la \E{\te{X\si_x}}-\frac{\ga+\iGa}{2}\E{\siy} \\
 \Dt{\E{\siz}}   & = \iDe \E{\siy}-\frac{\ga+\iGa}{2}\E{\siz}
\eea
The equations of motion for the first order moments couple to the second order moments. We define the semiclassical equations by factorising these second order moments to get a closed system of equations. Specifically, this means that we make the two assumptions that $\E{\te{X\six}}=\E{\te{X}}\E{\six}$ and $\E{\te{X\siy}}=\E{\te{X}}\E{\siy}$, or equivalently that $\E{\te{X},\six}\ll\E{\te{X}}\E{\six}$ and $\E{\te{X},\siy}\ll\E{\te{X}}\E{\siy}$. (An interesting special case arises when $\iDe=0$. Then equations of motion for $\E{\te{X}}$, $\E{\te{Y}}$, and $\E{\siz}$ are seen to decouple from those for $\E{\six}$ and $\E{\siy}$ and consequently higher order moments. This decoupled system of $\E{\te{X}}$, $\E{\te{Y}}$, and $\E{\siz}$, can be exactly solved). After factorising the second order moments, the semiclassical variables are defined by:
\bea
 \E{\te{X}} & \mapsto x \\
 \E{\te{Y}} & \mapsto y \\
 \E{\six}   & \mapsto L_x \\
 \E{\siy}   & \mapsto L_y \\
 \E{\siz}   & \mapsto L_z
 \label{e:scv}
\eea
Yielding the semi-classical equations of motion (where a dot indicates a time derivative):
\bea
 \dot{x}   & = \om y-\ka x \\
 \dot{y}   & = -\et-\la L_z-\om x-\ka y \\
 \dot{L_x} & = -\ep L_y-4\la xL_y-\ga\<{1+L_x} \\
 \dot{L_y} & = -\iDe L_z+\ep L_x+4\la xL_x-\frac{\ga+\iGa}{2}L_y \\
 \dot{L_z} & = \iDe L_y-\frac{\ga+\iGa}{2}L_z
 \label{e:eom}
\eea
whose steady states must satisfy the Bloch sphere constraints, which depend on the presence or absence of qubit dissipation. If there is no qubit dissipation, the steady states lie on the Bloch sphere; whereas with the presence of qubit dissipation they may lie inside:
\bea
 L_x^2+L_y^2+L_z^2 & =   1 && \textrm{ if }\ga=\iGa=0 \\
 L_x^2+L_y^2+L_z^2 & \le 1 && \textrm{ if }\ga>0\textrm{ or }\iGa>0
 \label{e:bs}
\eea

\subsection{Semi-classical fixed points locations}

The semi-classical equations of motion, \eqref{e:eom}, have fixed points ($\dot x=\dot y=\dot L_x=\dot L_y=\dot L_z=0$) that satisfy the Bloch sphere constraints, \eqref{e:bs}. These solutions are markedly qualitatively different depending on the presence or absence of qubit dissipation. In fact, there are three qualitatively different semi-classical steady states: no qubit dissipation; qubit dissipation consisting of only dephasing (no spontaneous emission); and qubit dissipation with spontaneous emission. The possible presence of oscillator/cavity decay is included in each of the three categories. The previous coupling positivities ($\om,\iDe,\la>0$ and $\ka,\ga,\iGa\ge0$ and $\ep,\et\in\as{R}$) will be assumed for all of the following semi-classical analysis.

We also define several convenient parameters: first, the bias / driving parameter $\xi$, which we see is zero in the case of an external voltage $v=-\frac{e}{C_M}$; second, a parameter $\al$ dependent on the coupling $\la$; third, a parameter $\be$ dependent on the magnitude of the qubit dissipation parameters; fourth, a parameter $\de$ dependent on the ratio between the two different types of qubit dissipation (spontaneous emission $\ga$ and dephasing $\iGa$); fifth, we also define two combinations of these parameters: $\mu$ and $\nu$. The previous assumptions about coupling positivities imply similar assumptions about these parameters ($\al,\mu>0$ and $\be,\de,\nu\ge1$ and $\xi\in\as{R}$).
\beq
 \xi = -\frac{\et}{\la}+\frac{\ep\<{\om^2+\ka^2}}{4\la^2\om}
\eeq
\bea
 \al & = \frac{4\la^2\om}{\iDe\<{\om^2+\ka^2}}, & 
 \be & = 1+\<{\frac{\ga+\iGa}{2\iDe}}^2,        &
 \de & = 1+\frac{\iGa}{\ga}
\eea
\bea
 \mu = \de\al & = \frac{4\<{1+\frac{\iGa}{\ga}}\la^2\om}{\iDe\<{\om^2+\ka^2}}, &
 \nu = \de\be & = \<{1+\frac{\iGa}{\ga}}\<{1+\<{\frac{\ga+\iGa}{2\iDe}}^2}
\eea

We must also consider the stability of the fixed points. The five semi-classical variables can be considered as a vector $\ma{x}$, such that about a fixed point $\ma{x^0}$ we have:
\beq
 \gma{\de}\ma{x} = \ma{x}-\ma{x^0} = \pb{x-x^0,y-y^0,L_x-L_x^0,L_y-L_y^0,L_z-L_z^0}^T
\eeq
\beq
 \Dt{}\gma{\de}\ma{x} = \ma{M}\gma{\de}\ma{x}
\eeq
where the Jacobian matrix $\ma{M}$ is:
\beq
 \ma{M} = \bm{
   -\ka        & \om  & 0            & 0                   & 0\\
   -\om        & -\ka & 0            & 0                   & -\la\\
   -4\la L_y^0 & 0    & -\ga         & -\ep-4\la x^0       & 0\\
   4\la L_x^0  & 0    & \ep+4\la x^0 & -\frac{\ga+\iGa}{2} & -\iDe\\
   0           & 0    & 0            & \iDe                & -\frac{\ga+\iGa}{2} }
\eeq
Stability of the fixed point requires all the eigenvalues of the Jacobian to have a real part less than or equal to zero. In general stability depends on many more coupling parameter combinations than those which define the fixed points. Thus, the stability is generally calculated numerically for fixed points with specific values of all couplings.

\subsubsection{No qubit dissipation}

When qubit dissipation is neglected ($\ga=\iGa=0$, $\ka\ge0$), there are three classes of semi-classical steady states. Two of these classes require $\xi=0$; and the other requires $\xi\ne0$.

\renewcommand{\thefootnote}{\fnsymbol{footnote}}

Class 1: for $\xi=0$ there are two fixed points that occur for all parameter values:
\beaa
 x^0   & = -\frac{\eta\om}{\om^2+\ka^2} = -\frac{\ep}{4\la} \\
 y^0   & = -\frac{\ka\eta}{\om^2+\ka^2} = -\frac{\ka\ep}{4\la\om} \\
 L_x^0 & = \pm1 \\
 L_y^0 & = 0 \\
 L_z^0 & = 0
\eeaa
The fixed point with $L_x^0=+1$ is unstable for almost all\footnotemark[1]\footnotetext[1]{Stability was determined numerically. This involved taking random samples in the parameter space spanned by $\om$, $\iDe$, $\ep$, $\la$, $\et$, $\ka$, $\ga$, and $\iGa$. ``Almost always'' stable/unstable indicates that the vast majority but not all ($>99\%$ but $<100\%$) of the random samples taken yielded stable/unstable fixed points.} coupling values; while the fixed point with $L_x^0=-1$ is almost always stable for $\al<1$ and almost always unstable for $\al>1$.

\begin{figure}[!htbp]\begin{center}
\includegraphics[scale=0.5]{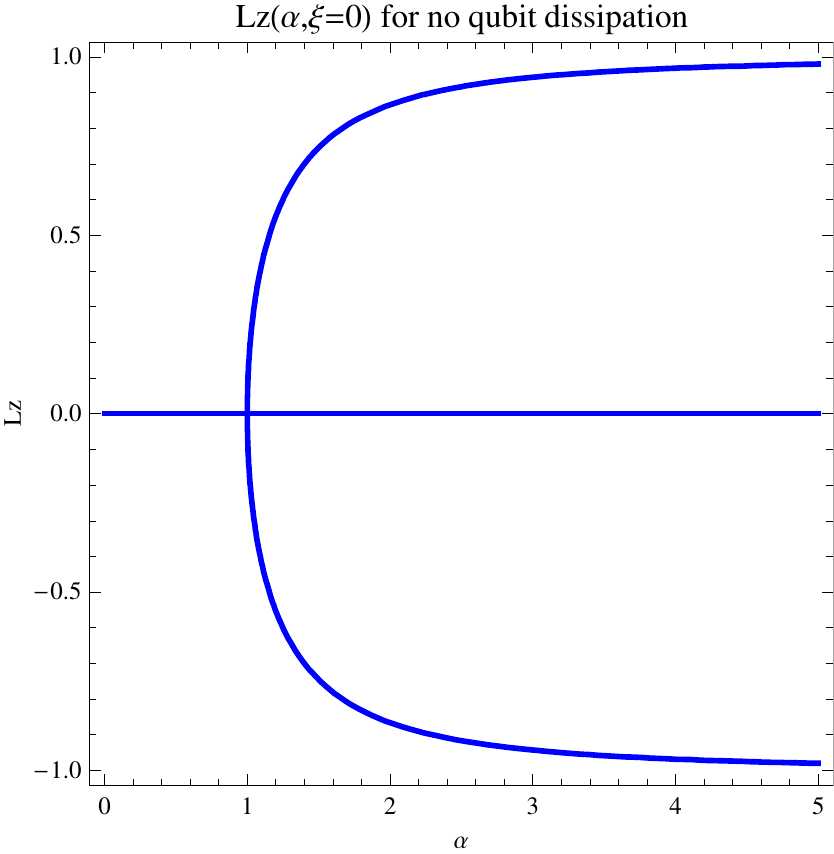}
\caption{$L_z$ component of the semi-classical steady states as a function of the parameter $\al$ for no spontaneous emission or dephasing of the two-level system ($\ga=\iGa=0$) and no driving ($\xi=0$). There are two solutions along the line $L_z^0=0$, one of which is almost always stable for $\al<1$; otherwise these $L_z^0=0$ solutions are almost always unstable. The two new solutions which appear for $\al>1$ have $L_z$ components $L_z^0=\pm\sqrt{1-\frac{1}{\al^2}}$ and can be stable or unstable depending on the coupling values (for example they are almost always stable when there is no oscillator decay). Hence, depending on the values of the coupling parameters there is often a supercritical pitchfork bifurcation at $\al=1$.}
\label{f:Lz_none_xi0}
\end{center}\end{figure}

Class 2: also for $\xi=0$ there are two fixed points that only occur for $\al>1$ (note that for $\al=1$ this second class of fixed points is also the first class just described):
\beaa
 x^0   & = -\frac{\eta\om}{\om^2+\ka^2}\mp\frac{\la\om}{\om^2+\ka^2}\sqrt{1-\frac{1}{\al^2}} && = -\frac{\ep}{4\la}\mp\frac{\la\om}{\om^2+\ka^2}\sqrt{1-\frac{1}{\al^2}} \\
 y^0   & = -\frac{\ka\eta}{\om^2+\ka^2}\mp\frac{\ka\la}{\om^2+\ka^2}\sqrt{1-\frac{1}{\al^2}} && = -\frac{\ka\ep}{4\la\om}\mp\frac{\ka\la}{\om^2+\ka^2}\sqrt{1-\frac{1}{\al^2}} \\
 L_x^0 & = -\frac{1}{\al} \\
 L_y^0 & = 0 \\
 L_z^0 & = \pm\sqrt{1-\frac{1}{\al^2}}
\eeaa
These fixed points can be stable or unstable depending on the coupling values (for example they are almost always ($>99\%$) stable for $\ka=0$). The $L_z$ components of the first two classes of fixed points are plotted in figure \ref{f:Lz_none_xi0}.

Class 3: for $\xi\ne0$ there are up to four real fixed points dependent on a quartic equation:
\bea
 x^0   & = -\frac{\eta\om}{\om^2+\ka^2}-\frac{\la\om}{\om^2+\ka^2}L_z^0 \\
 y^0   & = -\frac{\ka\eta}{\om^2+\ka^2}-\frac{\ka\la}{\om^2+\ka^2}L_z^0 \\
 L_x^0 & = -\frac{1}{\al}\frac{L_z^0}{L_z^0-\xi} \\
 L_y^0 & = 0 \\
 L_z^0 & = L_z^0
\eea
where $L_z^0$ satisfies the quartic equation (from \eqref{e:bs}):
\beq
 \label{e:quartic}
 \<{L_z^0}^2+\al^2\<{\<{L_z^0}^2-1}\<{L_z^0-\xi}^2=0
\eeq
Note that for $\xi\ne0$, $L_z^0=\xi$ is never a solution to this equation and so the pole in the expression for $L_x^0$ above is never encountered.

\begin{figure}[!htbp]\begin{center}
\includegraphics[scale=0.5]{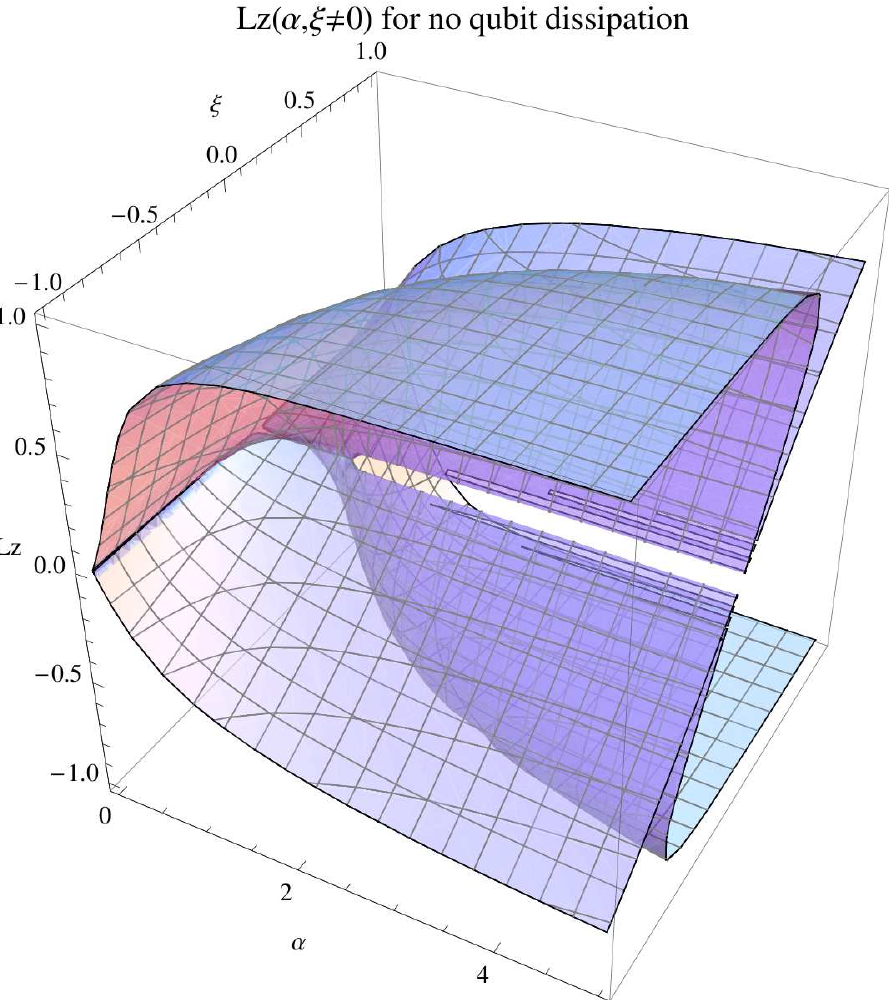}
\caption{$L_z$ component of the semi-classical steady states as a function of the parameters $\al$ and $\xi$ for no spontaneous emission or dephasing of the two-level system ($\ga=\iGa=0$) with driving ($\xi\ne0$). There are up to four solutions where the $L_z$ components are the real roots $L_z^0$ of the quartic equation $\<{L_z^0}^2+\al^2\<{\<{L_z^0}^2-1}\<{L_z^0-\xi}^2=0$. It is clear that varying either $\al$ or $\xi$ can take the solutions through bifurcations. This is shown explicitly in the bifurcation diagram of figure \ref{f:bif_none_xinot0}.}
\label{f:Lz_none_xinot0}
\end{center}\end{figure}

\begin{figure}[!htbp]\begin{center}
\includegraphics[scale=0.5]{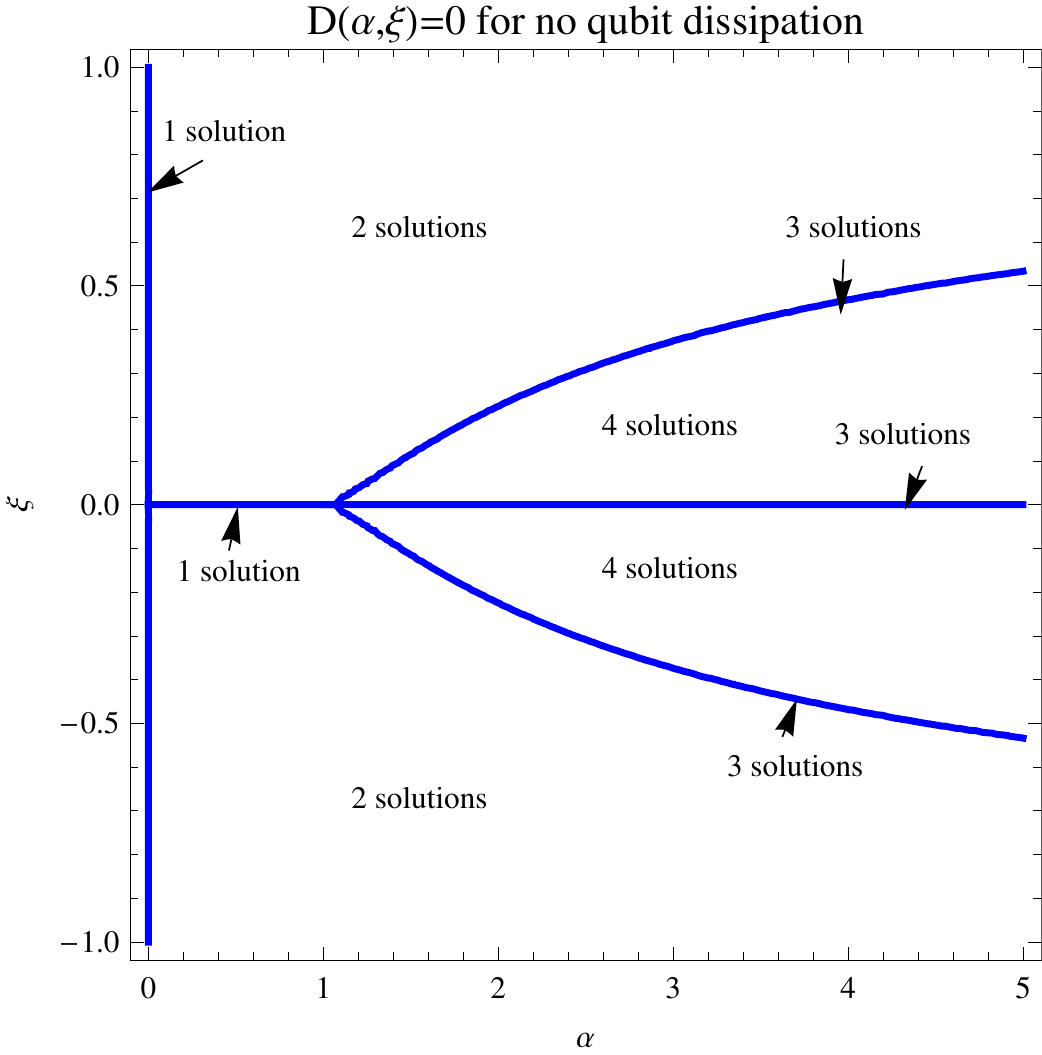}
\caption{Bifurcation diagram of the semi-classical steady states as a function of the parameters $\al$ and $\xi$ for no spontaneous emission or dephasing of the two-level system ($\ga=\iGa=0$) with driving ($\xi\ne0$). The bifurcations occur along the contours $\al^4\xi^2\<{\<{\xi^2-1}^3\al^6+3\<{\xi^4+7\xi^2+1}\al^4+3\<{\xi^2-1}\al^2+1}=0$. The extra dimension shows that an increase in the magnitude of the driving parameter $\xi$ means that a stronger coupling between the oscillator and two-level system is required to cross the bifurcation. The $L_z$ components of the fixed points are illustrated in figure \ref{f:Lz_none_xinot0} and the bifurcations are clearly visible. Also, the $\xi=0$ line correctly reflects the bifurcation shown for the no driving case in figure \ref{f:Lz_none_xi0}.}
\label{f:bif_none_xinot0}
\end{center}\end{figure}

The $L_z$ component of this third class of fixed points is plotted in figure \ref{f:Lz_none_xinot0}. The bifurcations of these fixed points are shown in the bifurcation diagram of figure \ref{f:bif_none_xinot0}.

\subsubsection{Dephasing-only qubit dissipation}

When the qubit dissipation is considered to consist of only phase decay ($\ga=0$, $\iGa>0$, $\ka\ge0$), there are two classes of semi-classical steady states. One of these classes requires $\xi=0$; and the other requires $\xi\ne0$.

Class 1: for $\xi=0$ there are infinite fixed points that occur for all parameter values:
\beaa
 x^0   & = -\frac{\eta\om}{\om^2+\ka^2} = -\frac{\ep}{4\la} \\
 y^0   & = -\frac{\ka\eta}{\om^2+\ka^2} = -\frac{\ka\ep}{4\la\om} \\
 L_x^0 & = L_x^0 \\
 L_y^0 & = 0 \\
 L_z^0 & = 0
\eeaa
where $\<{L_x^0}^2\le1$. These fixed points can be stable or unstable depending on the coupling values and the choice of $L_x^0$.

Class 2: for $\xi\ne0$ there is one (trivial) fixed point that occurs for all parameter values:
\bea
 x^0   & = -\frac{\eta\om}{\om^2+\ka^2} \\
 y^0   & = -\frac{\ka\eta}{\om^2+\ka^2} \\
 L_x^0 & = 0 \\
 L_y^0 & = 0 \\
 L_z^0 & = 0
\eea
This fixed point is stable for all coupling values.

\subsubsection{General qubit dissipation}

The general case of qubit dissipation here means with spontaneous emission present ($\ga>0,\iGa\ge0,\ka\ge0$), in which case there are three classes of semi-classical steady states. Two of these classes require $\xi=0$; and the other requires $\xi\ne0$.

Class 1: for $\xi=0$ there is one fixed point that occur for all parameter values:
\beaa
 x^0   & = -\frac{\eta\om}{\om^2+\ka^2} = -\frac{\ep}{4\la} \\
 y^0   & = -\frac{\ka\eta}{\om^2+\ka^2} = -\frac{\ka\ep}{4\la\om} \\
 L_x^0 & = -1 \\
 L_y^0 & = 0 \\
 L_z^0 & = 0
\eeaa
This fixed point is always stable for $\mu<\nu$ and always unstable for $\mu>\nu$.

Class 2: also for $\xi=0$ there are two fixed points that only occur for $\al>\be$ (note that for $\al=\be$ this second class of fixed points is also the first class just described):
\bea
 x^0   & = -\frac{\eta\om}{\om^2+\ka^2}\mp\sqrt{2}\frac{\la\om}{\om^2+\ka^2}\sqrt{\frac{\al-\be}{\de\al^2}} && = -\frac{\ep}{4\la}\mp\sqrt{2}\frac{\la\om}{\om^2+\ka^2}\sqrt{\frac{\al-\be}{\de\al^2}} \\
 y^0   & = -\frac{\ka\eta}{\om^2+\ka^2}\mp\sqrt{2}\frac{\ka\la}{\om^2+\ka^2}\sqrt{\frac{\al-\be}{\de\al^2}} && = -\frac{\ka\ep}{4\la\om}\mp\sqrt{2}\frac{\ka\la}{\om^2+\ka^2}\sqrt{\frac{\al-\be}{\de\al^2}}\\
 L_x^0 & = -\frac{\be}{\al} \\
 L_y^0 & = \pm\sqrt{2}\sqrt{\be-1}\sqrt{\frac{\al-\be}{\de\al^2}} \\
 L_z^0 & = \pm\sqrt{2}\sqrt{\frac{\al-\be}{\de\al^2}}
\eea
These fixed points can be stable or unstable depending on the coupling values.

\begin{figure}[!htbp]\begin{center}
\includegraphics[scale=0.5]{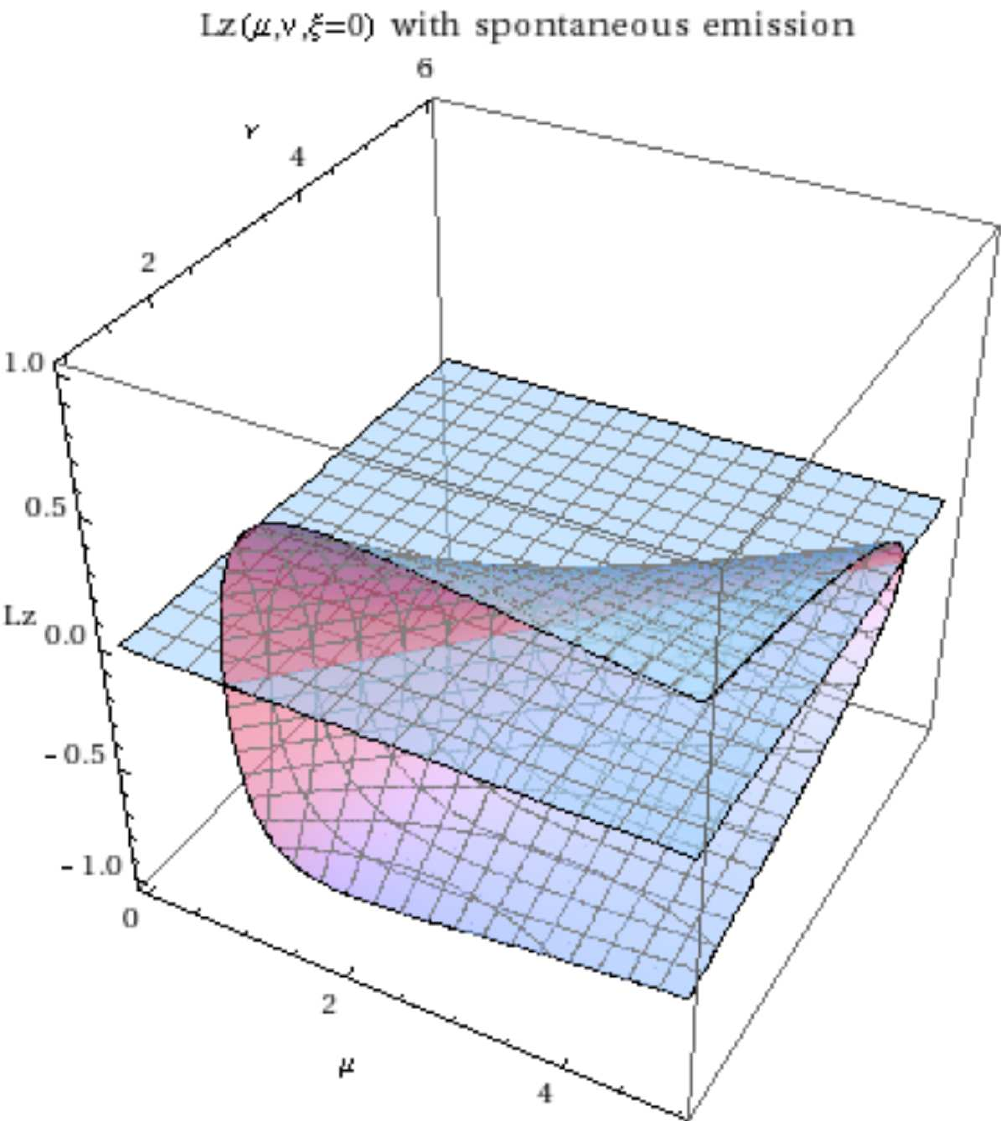}
\caption{$L_z$ component of the semi-classical steady states as a function of the parameters $\mu$ and $\nu$ when spontaneous emission is present ($\ga>0$) but there is no driving ($\xi=0$). There is one solution $L_z^0=0$ which is always stable for $\mu<\nu$ and always unstable for $\mu>\nu$. The two new solutions which appear for $\mu>\nu$ have $L_z$ components $L_z^0=\pm\sqrt{2}\sqrt{\frac{\al-\be}{\de\al^2}}$ and can be stable or unstable depending on the coupling values. Hence there is a pitchfork bifurcation along the line $\mu=\nu$ which is often supercritical depending on the coupling parameters. This bifurcation is shown explicitly in the bifurcation diagram of figure \ref{f:bif_se_xi0}.}
\label{f:Lz_se_xi0}
\end{center}\end{figure}

\begin{figure}[!htbp]\begin{center}
\includegraphics[scale=0.5]{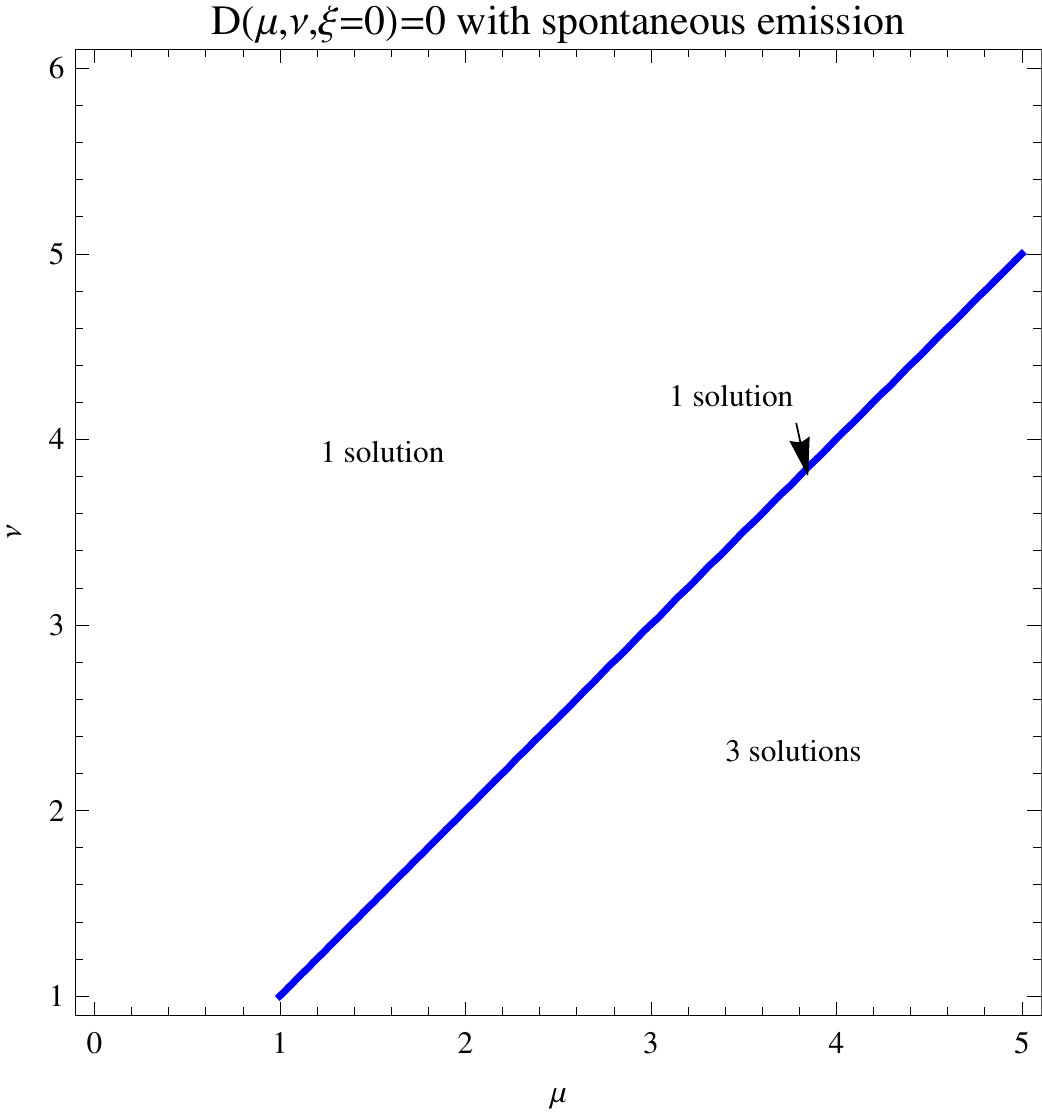}
\caption{Bifurcation diagram of the semi-classical steady states as a function of the parameters $\mu$ and $\nu$ when spontaneous emission is present ($\ga>0$) but there is no driving ($\xi=0$). The bifurcations occur along the line $\mu=\nu$. The extra dimension shows that an increase in the qubit dissipation parameter $\nu$ means that a stronger coupling between the oscillator and two-level system is required to cross the bifurcation. The $L_z$ components of the fixed points are illustrated in figure \ref{f:Lz_se_xi0} and the bifurcation is clearly visible.}
\label{f:bif_se_xi0}
\end{center}\end{figure}

The $L_z$ components of the first two classes of fixed points are plotted in figure \ref{f:Lz_se_xi0}. The bifurcations of these fixed points are shown in the bifurcation diagram of figure \ref{f:bif_se_xi0}.

Class 3: for $\xi\ne0$ there are up to three real fixed points dependent on a cubic equation:
\bea
 x^0   & = -\frac{\eta\om}{\om^2+\ka^2}-\frac{\la\om}{\om^2+\ka^2}L_z^0 \\
 y^0   & = -\frac{\ka\eta}{\om^2+\ka^2}-\frac{\ka\la}{\om^2+\ka^2}L_z^0 \\
 L_x^0 & = -\frac{\be}{\al}\frac{L_z^0}{L_z^0-\xi} \\
 L_y^0 & = \sqrt{\be-1}\;L_z^0 \\
 L_z^0 & = L_z^0
\eea
where $L_z^0$ satisfies the cubic equation:
\beq
 \label{e:cubic}
 \frac{1}{2}\mu^2L_z^0\<{L_z^0-\xi}^2-\mu\<{L_z^0-\xi}+\nu L_z^0=0
\eeq
Note that for $\xi\ne0$, $L_z^0=\xi$ is never a solution to this equation and so the pole in the expression for $L_x^0$ above is never encountered.

If we consider the third class of fixed points at the forbidden point $\xi=0$, this third class of fixed points gives the first (except for $L_x^0$) and second classes of fixed points; hence it generalises the first two classes in a sense.

\begin{figure}[!htbp]\begin{center}
\includegraphics[scale=0.5]{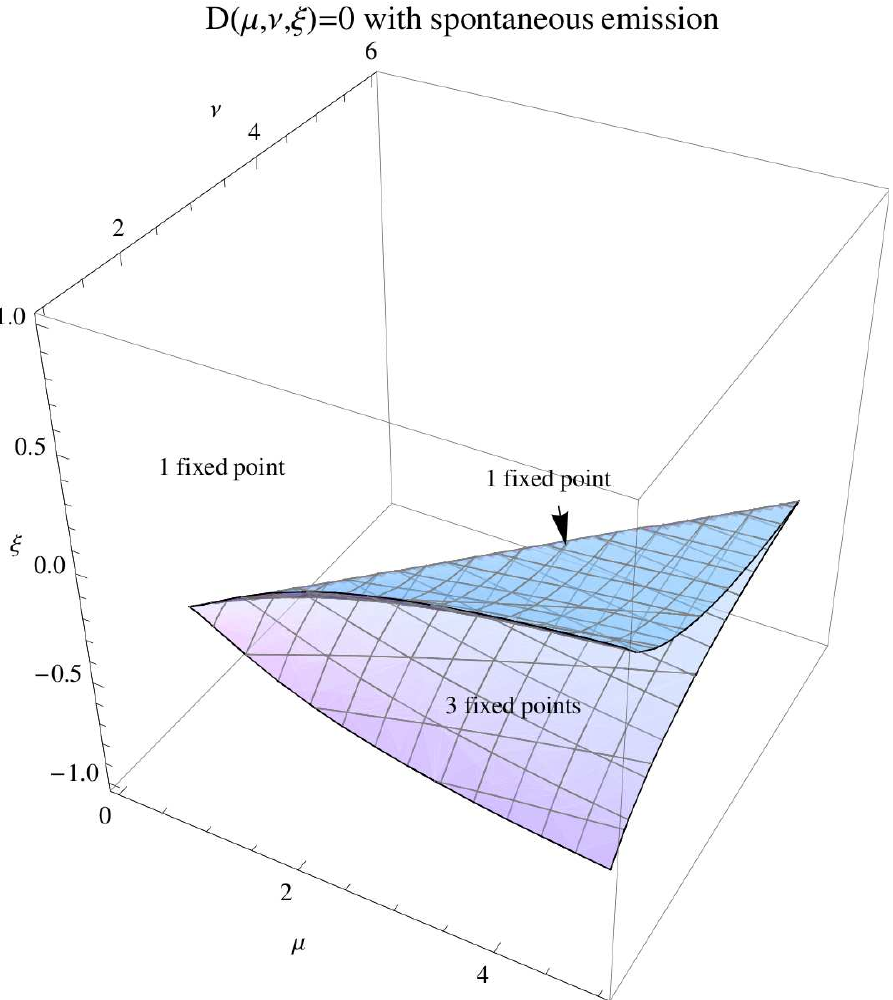}
\caption{Bifurcation diagram of the semi-classical steady states as a function of the parameters $\mu$ and $\nu$ when spontaneous emission is present ($\ga>0$) with driving ($\xi\ne0$). The bifurcations occur along the contours $-2\nu\xi^4\mu^6+\<{\mu^2-20\nu\mu-8\nu^2}\xi^2\mu^4+8\<{\mu-\nu}^3\mu^2=0$. The extra dimensions show that an increase in either the qubit dissipation parameter $\nu$ or the magnitude of the driving parameter $\xi$ means that a stronger coupling between the oscillator and two-level system is required to cross the bifurcation. Note that the $\xi=0$ cross-section correctly reflects the bifurcation shown for the no driving case in figure \ref{f:bif_se_xi0}.}
\label{f:bif_se_xinot0}
\end{center}\end{figure}

The $L_z$ component of this third class of fixed points is a function of the three parameters $\mu$, $\nu$,  and $\xi$. The bifurcations of these fixed points are shown in the bifurcation diagram of figure \ref{f:bif_se_xinot0}.

\section{Quantum steady states.}

Knowing the coupling parameter values that result in a semi-classical bifurcation of the steady state solutions, we wish to investigate whether there is a correspondence with the full quantum version. We do this numerically and observe the steady state phase space of the oscillator as we change the coupling parameters to move through the semi-classical fixed point bifurcation. It is hoped that our semi-classical analysis of the fixed points can be numerically justified by observing a signature of the semi-classical bifurcation.

\begin{figure}[!htbp]\begin{center}
\includegraphics[scale=0.5]{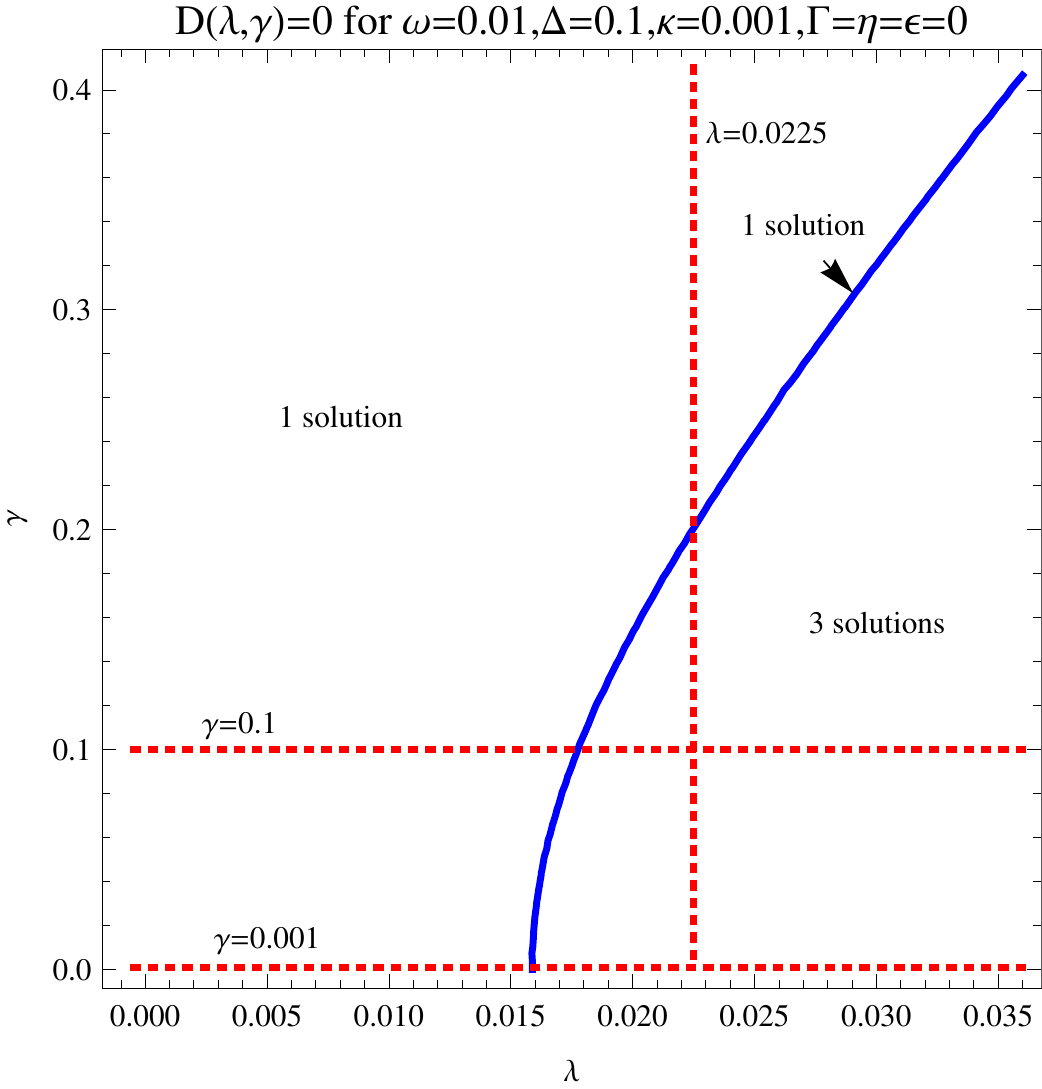}
\caption{Bifurcation diagram of the semi-classical steady states as a function of the parameters $\la$ and $\ga$ for the parameters used to numerically calculate the Q-functions: the first series of Q-functions in figure \ref{f:husimi_las_small_ga} is for $\la$ increasing along the lower horizontal dotted red line $\ga=0.001$; the second series of Q-functions in figure \ref{f:husimi_las_big_ga} is for $\la$ increasing along the upper horizontal dotted red line $\ga=0.1$; and the third series of Q-functions in figure \ref{f:husimi_gas} is for $\ga$ increasing along the vertical dotted red line $\la=0.0225$.}
\label{f:bif_matlab}
\end{center}\end{figure}

Here, we use the Quantum optics MATLAB toolbox \cite{MatlabQOT} and pass through two semi-classical bifurcations: one by varying the oscillator-qubit coupling ($\la$); and another by varying the spontaneous emission ($\ga$). By holding all other couplings equal and ignoring dephasing, varying these two parameters directly corresponds to varying the parameters $\mu$ and $\nu$ respectively. Specifically, we will look at a Jahn-Teller model where: $\om=0.01$, $\iDe=0.1$, $\ka=0.001$, $\iGa=0$, $\et=0$, and $\ep=0$. Thus the three parameters on which the semi-classical bifurcation depends become: $\mu=3960.4\la^2$, $\nu=1+25\ga^2$, and $\xi=0$. The contour in the bifurcation diagram of figure \ref{f:bif_se_xi0} can thus be redrawn as a function of $\la$ and $\ga$. This is done in figure \ref{f:bif_matlab}.

\begin{figure}[!htbp]\begin{center}
\includegraphics[scale=0.2]{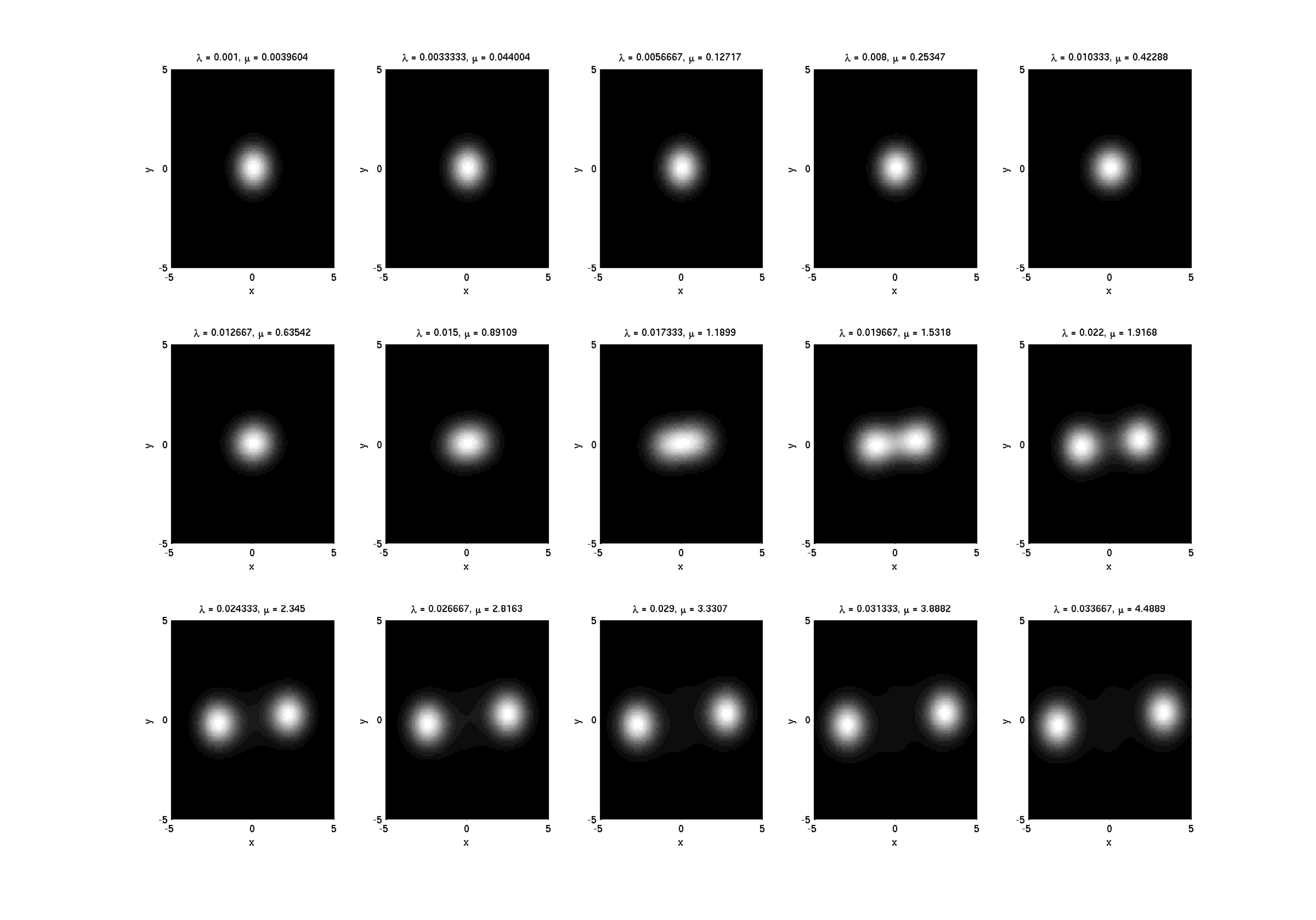}
\caption{Q-functions for increasing $\la$ along the lower horizontal dotted red line of \ref{f:bif_matlab}. The spontaneous emission is $\ga=0.0001$ giving a qubit dissipation parameter value of $\nu=1$. The semi-classical critical value of the oscillator-qubit coupling is $\la_c=0.01589$ at the critical parameter value $\mu_c=1$. The steady state phase space of the oscillator is seen to undergo a bifurcation which corresponds to the studied semi-classical bifurcation.}
\label{f:husimi_las_small_ga}
\end{center}\end{figure}

\begin{figure}[!htbp]\begin{center}
\includegraphics[scale=0.2]{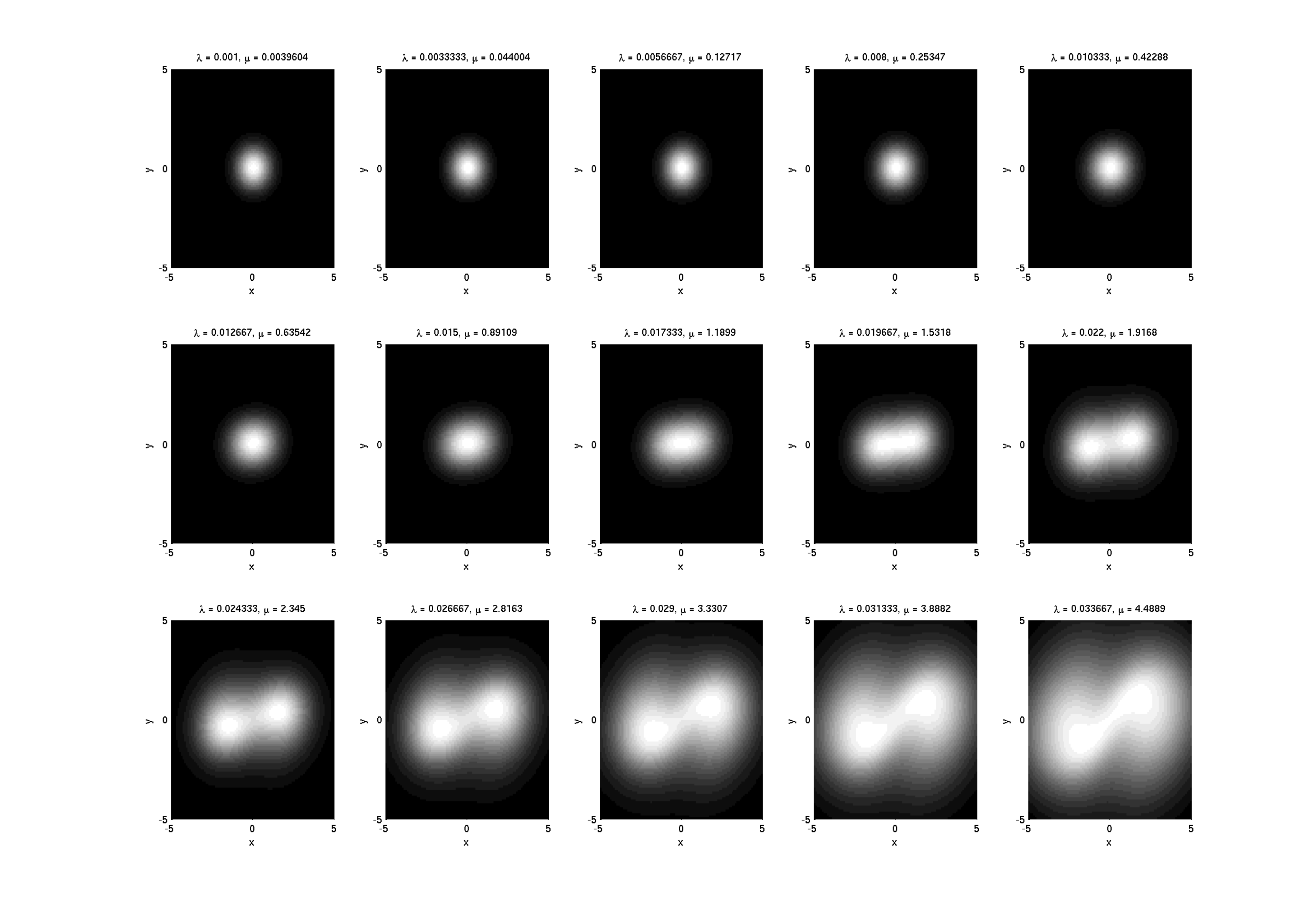}
\caption{Q-functions for increasing $\la$ along the upper horizontal dotted red line of \ref{f:bif_matlab}. The spontaneous emission is $\ga=0.1$ giving a qubit dissipation parameter value of $\nu=1.25$. The semi-classical critical value of the oscillator-qubit coupling is $\la_c=0.017766$ at the critical parameter value $\mu_c=1.25$. The steady state phase space of the oscillator is seen to undergo a bifurcation which corresponds to the studied semi-classical bifurcation.}
\label{f:husimi_las_big_ga}
\end{center}\end{figure}

\begin{figure}[!htbp]\begin{center}
\includegraphics[scale=0.2]{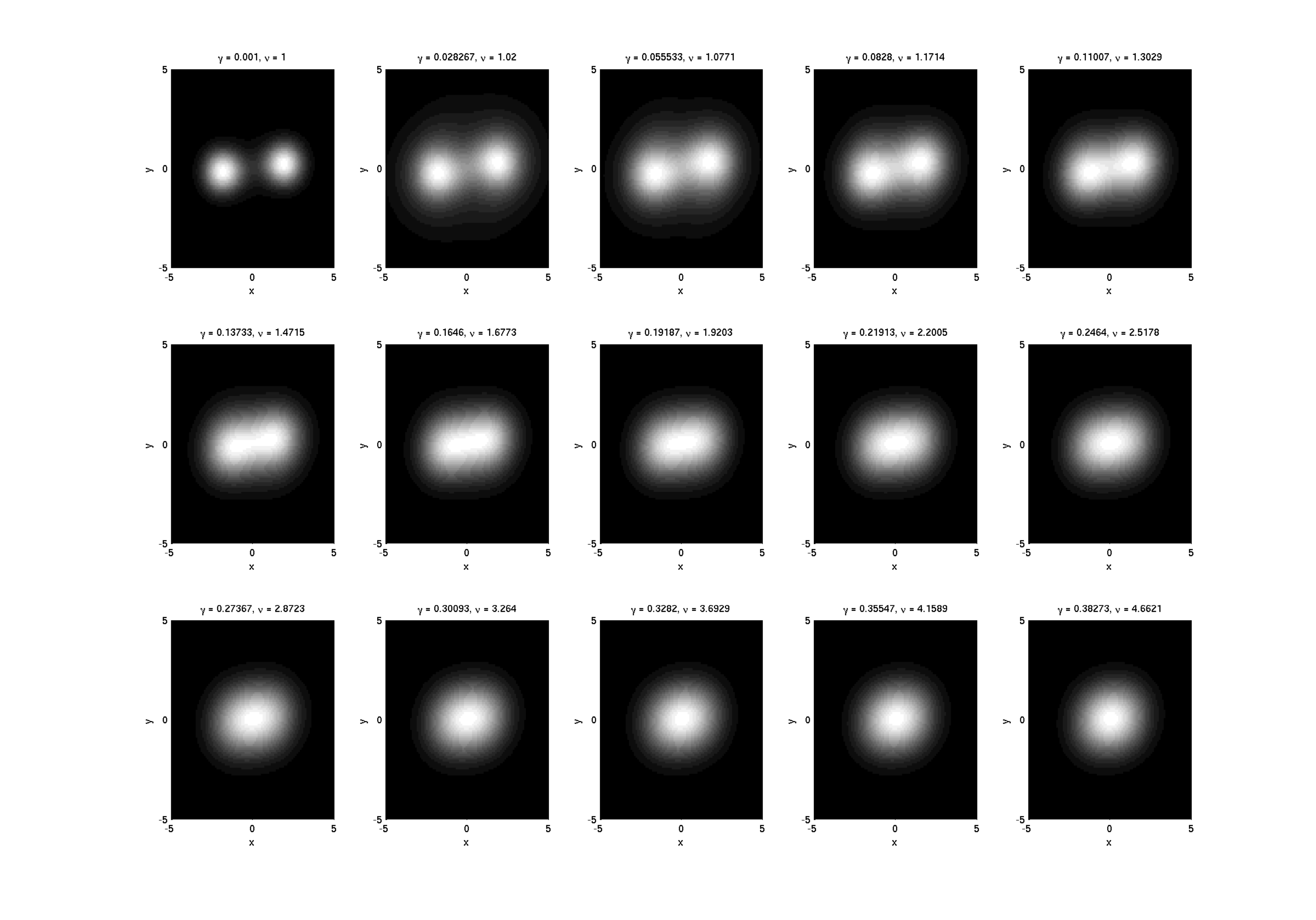}
\caption{Q-functions for increasing $\ga$ along the vertical dotted red line of Figure \ref{f:bif_matlab}. The oscillator-qubit coupling is $\la=0.0225$ giving a parameter value of $\mu=2.005$. The semi-classical critical value of spontaneous emission is $\ga_c=0.20049$ at the critical qubit dissipation parameter value $\nu_c=2.005$. The steady state phase space of the oscillator is seen to undergo a bifurcation which corresponds to the studied semi-classical bifurcation.}
\label{f:husimi_gas}
\end{center}\end{figure}

The MATLAB quantum optics toolbox gives us a steady state density matrix for the oscillator-qubit system. From this we can view the steady state phase space of the oscillator by plotting the Q-function for the corresponding reduced density operator of the oscillator. This is defined\cite{Walls-GJM} as the matrix elements of the reduced density operator for the oscillator in the coherent state basis, $Q(\alpha)= {\rm tr}(\rho|\alpha\rangle\langle\alpha|)$ where $|\alpha\rangle$ is a oscillator coherent state. 
Three series of Q-functions are plotted varying $\la$ for two differing fixed values of $\ga$, and varying $\ga$ for a fixed value of $\la$. These are shown in figures \ref{f:husimi_las_small_ga}, \ref{f:husimi_las_big_ga}, and \ref{f:husimi_gas} respectively. The semi-classical bifurcation is clearly evident in each case.

\section{Physical implementations of the Jahn-Teller model}

\subsection{A nanomechanical qubit.}
\begin{figure}[!htbp]
   \centering
   \includegraphics[scale=0.5]{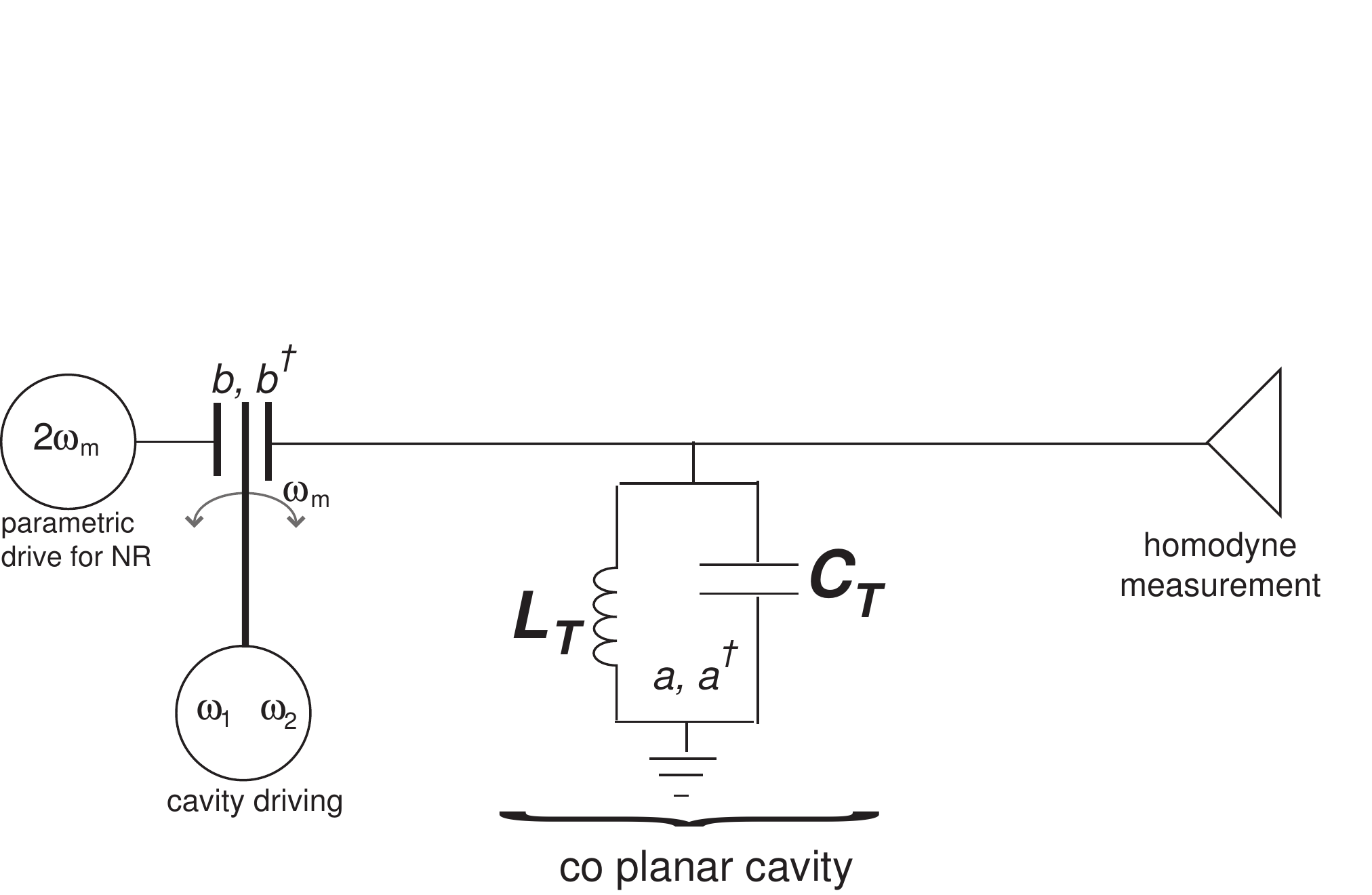} 
   \caption{Schematic of proposed nanomechanical qubit system.}
   \label{fig1}
\end{figure}

We consider a nonlinear nanomechanical resonator (NR), with resonant frequency $\omega_m$, coupled to a superconducting microwave resonator, with resonant frequency $\omega_c$, see figure \ref{fig1}. The NR is driven parametrically, while the cavity field is driven with two  microwave tones at frequencies $\omega_1$ and $\omega_2$   such that $(\omega_2-\omega_1)/2=\omega_m$ and with corresponding amplitudes $\epsilon_1,\epsilon_2$. 
The Hamiltonian is
\begin{eqnarray}
H & = & \hbar\omega_c a^\dagger a+\hbar\omega_m b^\dagger b+ \hbar\sum_{i=1}^2(\epsilon_i^*ae^{i\omega_i t}+\epsilon_i a^\dagger e^{-i\omega_i t})\\\nonumber
& & +\hbar\chi b^{\dagger 2} b^2+\hbar(\kappa^* b^2 e^{2i\omega_m t}+\kappa b^{\dagger 2} e^{-2i\omega_mt}) +\hbar g_0 a^\dagger a(b+b^\dagger)
\end{eqnarray}
where $a,a^\dagger$ are the lowering and raising operators for the cavity mode and $b,b^\dagger$ are the lowering and raising operators for the NR.  The term proportional to $\chi$ represents a quartic nonlinearity in the elastic potential energy of the NR and gives rise to a Duffing oscillator\cite{Babourina, Kozinsky}. The term proportional to $\kappa$ represents the parametric driving of the NR and has been discussed by \cite{Woolley1}. The term proportional to $g_0$ represents the capacitive coupling between the nanomechanical resonator and the cavity field, expanded to linear order in the displacement of the nanomechanical resonator.

We now move to an interaction picture for the NR at frequency $\omega_m$ and for the microwave cavity at frequency $(\omega_1+\omega_2)/2$. The Hamiltonian then becomes, 
\begin{eqnarray}
H & = & \hbar\omega a^\dagger a+ \hbar\sum_{i=1}^2(\epsilon_i^*ae^{i\delta_i t}+\epsilon_i a^\dagger e^{-i\delta_i t})\\\nonumber
& & +\hbar\chi b^{\dagger 2} b^2+\hbar(\kappa^* b^2+\kappa b^{\dagger 2}) +\hbar g_0 a^\dagger a(b e^{-i\omega_m t}+b^\dagger e^{i\omega_m t})
\end{eqnarray}
where the detuning of the cavity resonance is $\omega=\omega_c-(\omega_1+\omega_2)/2$, and $\delta_1=(\omega_1-\omega_2)/2=-\delta_2$. Following Woolley et al. \cite{Woolley1} we linearise around the steady state amplitudes for the cavity field and choose $\delta_1=-\omega_m$, with $\epsilon_1=-\epsilon_2^*=\epsilon e^{-i\psi}$ and we find the effective Hamiltonian
\begin{equation}
H_{e}= \hbar\omega a^\dagger a+\hbar g(a+a^\dagger)(be^{-i\psi}+b^\dagger e^{i\psi})+H_{NR}
\end{equation}
where
\begin{equation}
H_{NR}=\hbar\chi b^{\dagger 2} b^2+\hbar\kappa( b^2+b^{\dagger 2}) 
\end{equation}
with $\kappa$ real, and $g=g_0\epsilon/\omega_m$, and where  $\mu$ is the decay rate for the cavity field. Note that the coupling constant, $g$, can be made large by increasing the driving field $\epsilon$. 

The dynamics arising from $H_{NR}$ was investigated in \cite{Wielinga}. The classical phase space dynamics has two elliptic fixed points either side of a hyperbolic unstable fixed point.  This is equivalent to an effective double well potential. The quantum ground state is then seen to be very well approximated by a symmetric superposition of two oscillator coherent states $|\pm\alpha\rangle$ centered on the fixed points with
\begin{equation}
\alpha=-i\sqrt{\frac{\kappa}{\chi}}
\end{equation}
The energy separation between the ground state and first excited state was shown in \cite{Wielinga} to be 
\begin{equation}
\Delta E=\hbar \chi e^{-2|\alpha|^2}\equiv\hbar\Delta_s
\end{equation}
We now assume that the NR is always very close to its ground state and we truncate the Hilbert space to the ground state and the first excited state. We then define a qubit basis by
\begin{eqnarray}
|0\rangle & = & |\alpha\rangle \\\nonumber
|1\rangle & = & |-\alpha\rangle
\end{eqnarray}
It is then clear that $b|0\rangle=\alpha|0\rangle,\ b|1\rangle=-\alpha|1\rangle$. 
If we then define $\sigma_z=|1\rangle\langle 1|-|0\rangle\langle 0|,\ \sigma_x=|1\rangle\langle 0|+|0\rangle\langle 1|$, we can then write
\begin{equation}
H_{NR}=\hbar\frac{\Delta}{2}\sigma_x
\end{equation}
and the effective Hamiltonian, with the phase choice $\psi=\pi/2$, takes the form
\begin{equation}
H_{e}=\hbar\omega a^\dagger a+\hbar\frac{\Delta}{2}\sigma_x+\hbar\lambda (a+a^\dagger)\sigma_z
\end{equation}
where 
\begin{eqnarray}
\omega & = &  \omega_c-\frac{(\omega_1+\omega_2)}{2}\\
\lambda & = & 2g|\alpha|\\
\Delta & = & 2\chi e^{-2|\alpha|^2}
\end{eqnarray}
This is the Jahn-Teller model with a critical coupling strength $\lambda_{cr}$ given by\cite{JT-bifur}
\begin{equation}
\lambda_{cr}=\frac{\sqrt{\omega\Delta}}{2}
\end{equation}
If we add a resonant driving term to the NR, we get an additional term proportional to $\sigma_z$.

In \cite{Babourina}, the elastic nonlinearity of a Pt nanowire, implemented by Kozinsky et al.\cite{Kozinsky},  was estimated to have $\chi=10^{-4}\mbox{s}^{-1}$.  If we keep the parametric driving weak, we can ensure that $|\alpha|=5$ (see \cite{Woolley1}), this is a deep quantum regime in which the fixed points are separated from the unstable fixed point at the origin by a few units of ground state uncertainty. If we accept a detuning of about $10$MHz,  $\omega\sim 10^8$, then $\lambda_{cr}\sim 0.7\mbox{s}^{-1}$.  The coupling constant is largely determined by $g$ which depends on the intra-cavity mean field amplitude. This quantity can thus be controlled quite well.   Typical values \cite{Woolley1} for $g$ are of the order of $10-1000$s$^{-1}$, with the smaller number for bad cavities,  so it would be relatively easy to exceed the critical coupling strength.  


\subsection{Circuit QED}

\begin{figure}[!ht]\begin{center}
\includegraphics[scale=0.3]{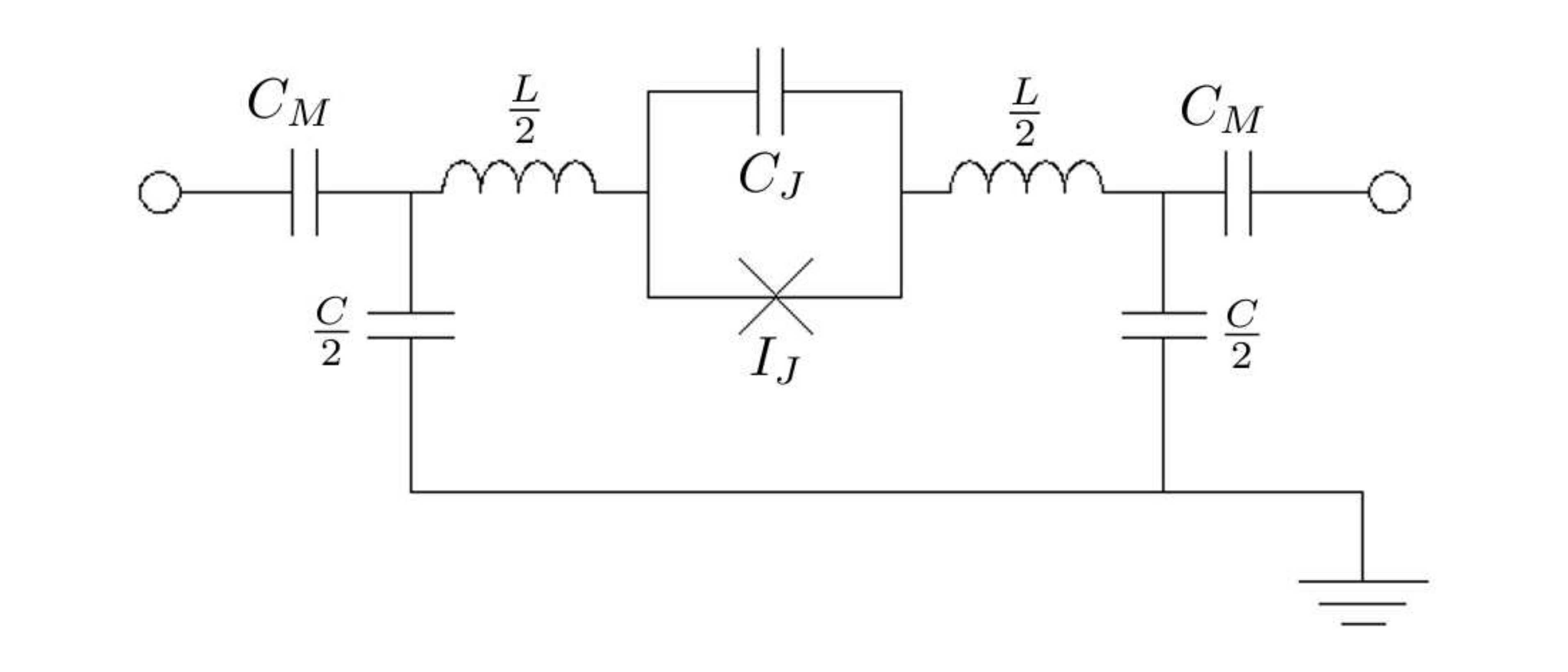}
\caption{Lumped element schematic circuit approximation.}
\label{devoret-circuit}
\end{center}\end{figure}

Devoret et al. \cite{Devoret:2007} have proposed a scheme to get ultra strong coupling between a Cooper pair box qubit and the microwave field of a coplanar superconducting resonator. The central conductor of the coplanar cavity is divided into two segments separated by a Cooper pair box, see figure \ref{devoret-circuit}. The quantum theory of such a system begins by first writing down the classical circuit dynamics, constructing a Lagrangian and an assocaited Hamiltonian. Quantistation then proceeds via the usual canonical method. This results in an {\em effective} quantum theory in which collective variables of direct interest to the experimentalist couple only weakly to the microscopic degrees of freedom, which remain as a source of dissipation and decoherence. 

\begin{figure}[!ht]\begin{center}
\includegraphics[scale=0.3]{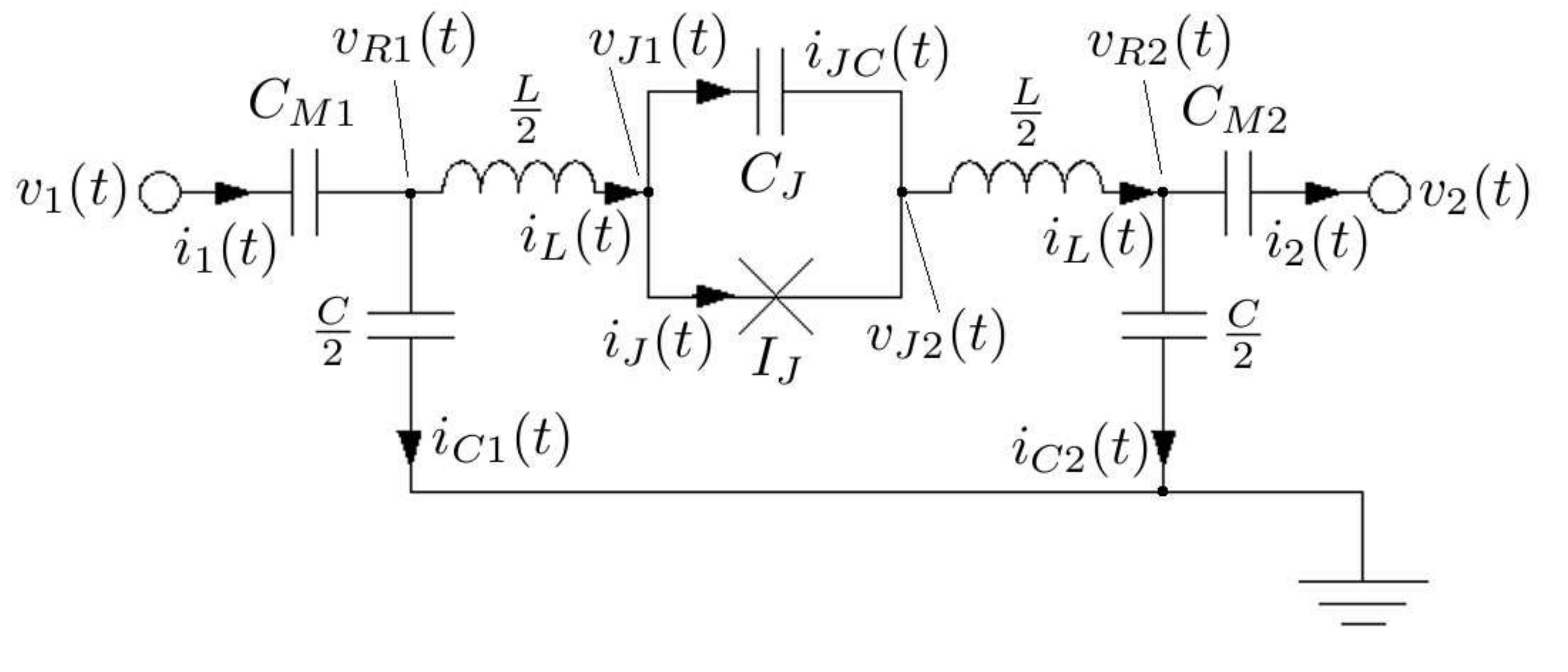}
\caption{Lumped element schematic circuit approximation, here labelled with the dynamic electrical variables.}
\label{f:devoret-circuit-labelled}
\end{center}\end{figure}

We label the relevant electrical variables as shown in figure \ref{f:devoret-circuit-labelled}. We consider an input and output currents $i_1(t)$ and $i_2(t)$ at voltages $v_1(t)$ and $v_2(t)$ respectively. The ``mirror'' capacitors of capacitance $C_{M1}$ and $C_{M2}$ hold charges $Q_{M1}(t)$ and $Q_{M2}(t)$ respectively and enable the device to be inserted into a transmission line. The lumped inductances $\f[L]{2}$ of the ``cavity'' hold magnetic fluxes $\Ph_1(t)$ and $\Ph_2(t)$. The lumped capacitances $\f[C]{2}$ of the cavity hold charges $Q_1(t)$ and $Q_2(t)$. The pure Josephson element, with a critical current $I_J$ and tunneling energy $E_J$, has a wavefunction phase difference across it of $\thet(t)$ and sees $N(t)$ Cooper pairs tunnel across it. The Josephson junction capacitance $C_J$ holds a charge $Q_{JC}(t)$. The currents flowing through the inductive and capacitive elements of the cavity are $i_L(t)$ and $i_{C1}(t)$ and $i_{C2}(t)$ respectively. The currents flowing through the pure Josephson element and the Josephson junction capacitance are $i_J(t)$ are $i_{JC}(t)$ respectively. Finally, the voltage at the input and output ends of the resonator are $v_{R1}(t)$ and $v_{R2}(t)$ respectively; and the voltage at the input and output ends of the Josephson junction are $v_{J1}(t)$ and $v_{J2}(t)$ respectively. Figure \ref{f:devoret-circuit-labelled} summarises all of this.

If we introduce creation and annihilation operators of the cavity field (the effective flux $\te{\Ph}$ acts as a position, and the effective charge $\te{Q}$ as a momentum):
\bea
 \te{\Ph} & = \ii\sqrt{\frac{\hb\om L}{2}}\<{\aop-\ad} \\
 \te{Q}   & = \sqrt{\frac{\hb\om C_E}{2}}\<{\aop+\ad}
\eea
where:
\bea
 \frac{1}{C_E} & = \frac{1}{C+C_M}+\frac{1}{C_J} \\
 \om           & = \frac{1}{\sqrt{LC_E}}
\eea
and if we consider only the bottom two energy levels for the Josephson junction using Pauli matrices for the resulting qubit, then we can rewrite the Hamiltonian (ignoring constant energy offsets) in the Jahn-Teller form of Eq.(\ref{JT-ham}) where, following Devoret et al.\cite{Devoret:2007}, 
\begin{equation}
\breve{g}\equiv\frac{2\lambda}{\Delta}=\frac{1}{\sqrt{8\pi}}\left (\frac{E_{C}}{2E_J}\right )^{1/4}\sqrt{\frac{Z_{vac}}{Z_C}}\alpha^{-1/2}
\end{equation}
with
\bea
 Z_{vac} & = \frac{1}{c\epsilon_0}\approx 377 \Omega\\
 Z_C     & = \sqrt{\frac{L}{C}}
\eea
Typically $Z_C=50\ \Omega$. Taking $E_C/E_J=200$, Devoret et al.\cite{Devoret:2007} arrive at a value of $\lambda\approx 10\Delta$ which is certainly well outside of the domain of validity for the rotating wave approximation. We thus believe this configuration offers a good chance of designing a system with a coupling strength above the Jahn-Teller dissispative bifurcation in circuit QED.

\section{Conclusion.}
In this paper we have presented a detailed analysis of the effect of dissipation on the dynamical bifurcation that occurs when there is strong coupling between a single two level system and an oscillator; a Jahn-Teller model. We have based our description of dissipation on the physically appropriate mechanisms two physical realisations of the system based on nanomechanicas and circuit Quantum electrodynamics. The key feature of the  bifurcation in the dissipative Jahn-Teller model is the change in the oscillator fixed point from one centered on a point of zero radius in phase-space  to one with support on a non zero value of the radius. This is a distinct kind of bifurcation from that discussed recently in the damped nanomechanical Duffing oscillator\cite{Kozinsky} which only involved a single degree of freedom. In the case considered here, the bifurcation results in steady state correlations between the state of the oscillaator and the two-level system.  

As the average excitation energy an oscillator is proportional to the radius in phase-space, this bifurcation would be reflected in a change in the steady state mean excitation energy from zero to a finite non zero value.  This would have implications for any attempt to cool the system though tuning to the red sideband transition, that is to say,  tuning the cavity field driving by the mechanical frequency below the cavity frequency. If the parameters were such that the system was already beyond the Jahn-Teller bifurcation, the mechanical system could not be cooled to a zero phonon state, but would rather relax to the bistable state with a non zero mean phonon number.  Fluctuations would then drive switching events between the two stable steady states. 

The non-dissipative model, for coupling stronger than the critical coupling, has a ground state with significant entanglement between the two-level system and the oscillator. We do not know if any entanglement remains in the steady state of the dissipative model beyond the bifurcation point. This is a difficult question to answer as the steady state has a non Gaussian Q-function (or Wigner function) and thus it is not clear what would be a good measure of entanglement. In a future work we will use positive P-function methods to attempt to answer this question. 

In many implementations of quantum information processing there is often an unwanted strong coupling between an oscillator degree of freedom and a strongly damped  two-level system\cite{chu,Deslauriers}. The model of this paper may be relevant to the on going study of such systems in those cases where a perturbative treatment of the coupling is not possible.

We have given two examples of quantum electromechanical systems that could exhibit the steady state bifurcation of the dissipative Jahn-Teller model. Observation of this effect would be a clear demonstration of the ultra strong coupling regime that can be achieved in these systems, as opposed to what typically happens in atomic systems where the rotating wave approximation eliminates the bifurcation. Such system open a path to study the quantum signature of non linear bifurcations on the steady states of strongly coupled systems. 

\acknowledgments
This work has been supported by the Australian Research Council. We would also like to thank Per Delsing and Andreas Wallraff for useful discussions.

\end{document}